\documentclass{ws-ijmpe}
\usepackage[super,compress]{cite}
\usepackage{lineno,hyperref}
\usepackage{amsmath}
\usepackage{supertabular}
\usepackage{multirow}
\usepackage{amssymb}
\usepackage{amsmath}
\usepackage{graphicx}
\usepackage[normalem]{ulem}
\usepackage{longtable}
\usepackage{bm}
\usepackage{array}
\usepackage{etoolbox}
\usepackage{setspace}
\usepackage{color}
\usepackage{ulem}
\usepackage{graphicx}
\usepackage{booktabs}
\usepackage{amssymb,bm,mathrsfs,bbm,amscd}

\begin{document}

\markboth{Authors' Names}{Instructions for typing manuscripts (paper's title)}

\catchline{}{}{}{}{}

\title{Systematic study of two-proton radioactivity half-lives based on a modified Gamow-like model}

\author{Hong-Ming Liu}

\address{School of Nuclear Science and Technology, University of South China, 421001 Hengyang, People's Republic of China}

\author{You-Tian Zou}
\address{School of Nuclear Science and Technology, University of South China, 421001 Hengyang, People's Republic of China}

\author{Xiao Pan}
\address{School of Nuclear Science and Technology, University of South China, 421001 Hengyang, People's Republic of China}

\author{Biao He}
\address{College of Physics and Electronics, Central South University, 410083 Changsha , People's Republic of China}

\author{Xiao-Hua Li}
\address{{School of Nuclear Science and Technology, University of South China, 421001 Hengyang, People's Republic of China}\\
lixiaohuaphysics@126.com}

\maketitle

\begin{history}

\end{history}

\begin{abstract}
 In the present work, we systematically study the two-proton ($2p$) radioactivity half-lives of  nuclei close to the proton drip line within a modified Gamow-like model. Using this model, the calculated $2p$ radioactivity half-lives can well reproduce the experimental data. Moreover, we use this model to predict the $2p$ radioactivity half-lives of  22 candidates whose $2p$ radioactivity is energetically allowed or observed but not yet quantied in evaluated nuclear properties table NUBASE2016. The predicted results are in good agreement with the ones obtained by using Gamow-like model, effective liquid drop model (ELDM), generalized liquid drop model (GLDM) as well as a four-parameter formula. 
\end{abstract}

\keywords{two-proton radioactivity;  modified Gamow-like model; half-life.}



\section{INTRODUCTION}	
In recent years, extensive studies of the ground-state masses and decay modes for proton-rich nuclei in the vicinity of the proton drip line  have been attracted much interest and became significant topics in modern nuclear physics field since they are conducive to understand the nuclear forces and isospin-symmetry-breaking effects\cite{Olsen2013,Blank2008,Cole1996,Ormand1996,Grigorenko2000,Soylu2021,Pan2021}. For odd-$Z$ nuclei between $Z$ = 51 and $Z$ = 83, proton radioactivity is the predominant decay mode, which was firstly observed in an isomeric state of $^{53}$Co in 1970\cite{Jackson1970,Cerny1970}. To date, there are about 44 proton radioactivity nuclei observed decay from ground states or low-lying isomeric states to ground states. However, for even-$Z$ nuclei,  the two-proton ($2p$) radioactivity phenomenon may occur. This new exotic decay mode was successively predicted by Zel'dovich and Goldansky in the begining of 1960\cite{Zel'dovich1960,Goldansky1960,Goldansky1961}. Subsequently, with the continuous progress of experimental techniques, the not true $2p$ radioactivity ($Q_{p}>0$ and  $Q_{2p}>0$, where $Q_{p}$ and  $Q_{2p}$ are the proton radioactivity and $2p$ radioactivity released energy, respectively) nuclei $i. e. ,$ $^{6}\rm{Be}$, $^{12}\rm{O}$ and $^{16}\rm{Ne}$ produced by short-lived nuclear resonances were observed\cite{Whaling1966,KeKelis1978,Woodward1983,Kryger1995,Suzuki2009,Jager2012}. In 2002, the true $2p$ radioactivity ( $Q_{p}<0$ and $Q_{2p}>0$ ) was firstly discovered in the decay of $^{45}$Fe in two independent experiments at GANIL and GSI, respectively\cite{Giovinazzo2002,Pfutzner2002}.  Later on, the $2p$ radioactivity phenomena of $^{19}$Mg, $^{48}$Ni, $^{54}$Zn and $^{67}$Kr were also reported in different experiments\cite{Blank2005,Dossat2005,Mukha2007,Goigoux2016}. 

For $2p$ radioactivity process, two protons may be simultaneously emitted from the mother nucleus in virtue of the proton pairing interaction and the odd-even binding energy effect. Based on this hypothesis, masses of models and/or formulas have been proposed to deal with this decay mode\cite{Grigorenko2001,Grigorenko2003,Sreeja2019,Brown2019,Cole1999,Cole1997,Cole1998,Brown2002,Goncalves2017,Delion2013,Tavares2018,Wang2021}. The calculated results obtained by using all of these methods can reproduce the experimental $2p$ radioactivity half-lives. In 2013, based on the Wentzel-Kramers-Brillouin (WKB) theory, Zdeb $et\ al.$ proposed a single parameter model named as Gamow-like model to study $\alpha$ decay and cluster radioactivity\cite{Zdeb2013}. Later, this model was extended to investigating proton radioactivity and $2p$ radioactivity\cite{Zdeb2016,Liu2021}. In this model, the inner potential is expressed as a square potential well and the outer one defaults to the Coloumb potential under the assumption of uniform charge distribution. Whereas, as a result of the inhomogeneous charge distribution of the nucleus, the charge superposition of the emitted particle and other factors, the electrostatic shielding effect should be consisdered in the outer potential. 

Recently, introducing an exponential-type electrostatic potential $i. e. ,$ Hulth$\acute{\rm{e}}$n potential to describe the outer potential, R. Budaca $et \ al.$ put forward an analytical model to calculate the proton radioactivity half-lives of nuclei with  51 $< Z <$ 83\cite{Budaca2017}. In this model, the only  one parameter $a$ expresses the electrostatic screening effect on Coloumb potential, which can be explained as short-range effects like charge diffuseness, inhomogeneous charge distribution, charge superposition, etc. Very recently, the same success was achieved with respect to $\alpha$ decay and $2p$ radioactivity \cite{Budaca2020,ZouCPC}. In our previous works, consisdering the electrostatic shielding effect, we modified the Gamow-like model and used this model to systematically study the half-lives of $\alpha$ decay and proton radioactivity nuclei\cite{Cheng2019,Chen2019}. Our calculated results are in good agreement with the experimental data. Consisdering the two protons emitted from parent nucleus in $2p$ radioactivity process being a quasi-bound $^{2}$He-like cluster penetrating barrier, this process maybe share the same theory with $\alpha$, cluster and proton radioactivity processes\cite{Buck1994,Zou2021,Ghodsi2020,Balasubramaniam2005,Moghaddari2020}. Whether this modified Gamow-like model can be used to study $2p$ radioactivity or not is certainly of interest. For this propose, in the present work we systematically analyze the half-lives of 2$p$ radioactive nuclei close to the proton drip line by using modified Gamow-like model.

This article is organized as follows. In next section, the theoretical framework of the modified Gamow-like model is briefly presented. The detailed results and discussion are presented in Section 3. Finally, a summary is given in Section 4.

\section{THEORETICAL FRAMEWORK}
The half-life of a $2p$ radioactivity nucleus is expressed as
\begin{equation}
T_{1/2} = \frac{ln2}{\lambda},
\label{subeq:1}
\end{equation}
where $\lambda$ denotes the decay constant, it can be given by
\begin{equation}
\lambda = S_{2p} \nu \emph{P}.
\label{subeq:2}
\end{equation}
Here $S_{2p} = G^2[A/ (A -2)]^{2n}\chi ^2$ denotes the two-proton preformation probability on parent nucleus surface with $G^2 = (2n)! / [2^{2n}(n!)^2]$ \cite{Anyas1974}, $n \approx(3Z)^{1/3} - 1$ is the average principal proton oscillator quantum number with $Z$ being the proton number of parent nucleus\cite{Bohr1969}. $A$ denotes the mass number of parent nucleus. $\chi ^2$ = 0.0143 obtained by fitting the experimental data of $^{45}$Fe, $^{19}$Mg, $^{48}$Ni and $^{54}$Zn\cite{Cui2020}. $\nu$ represents the collision frequency of the emitted two protons on the potential barrier. It can be obtained by the harmonic oscillator frequency present in the Nilsson potential \cite{Nilsson1955}
\begin{equation}
h \nu = \hbar \omega \simeq \frac{41}{A^{1/3}},
\label{subeq:3}
\end{equation}
where $h$, $\hbar$ and $\omega$ are the Planck constant, the reduced Planck constant and the angular frequency, respectively. 

$P$, the Gamow penetrability factor through the barrier, can be calculated by using the semi--classical Wentzel--Kramers--Brillouin (WKB) approximation:
\begin{equation}
P= \rm{exp}\left[-\frac{2}{\hbar}\int_{\emph{R}_{in}}^{\emph{R}_{out}} \sqrt{{2 \mu}({\emph{V}(\emph{r})-\emph{E}_\emph{k}})}\,d\emph{r}\right].
\label{subeq:4} 
\end{equation}
 Here $\emph{R}_{\rm{out}}$ is the classical outer turning point, which satisfies the condition $V(R_{\rm{out}}) = E_k$ with $E_k$ being the kinetic energy of emitted two protons. $\emph{R}_{\rm{in}} = r_{0} (A_{2p}^{1/3} + A_{d}^{1/3})$ represents the spherical square well radius, where $r_{0}$ is effective nuclear radius parameter, $A_{2p}$ and $A_{d}$ are the mass numbers of emitted two protons and daughter nucleus, respectively. 
 
\begin{figure}
\centering
\includegraphics[width=15.0cm]{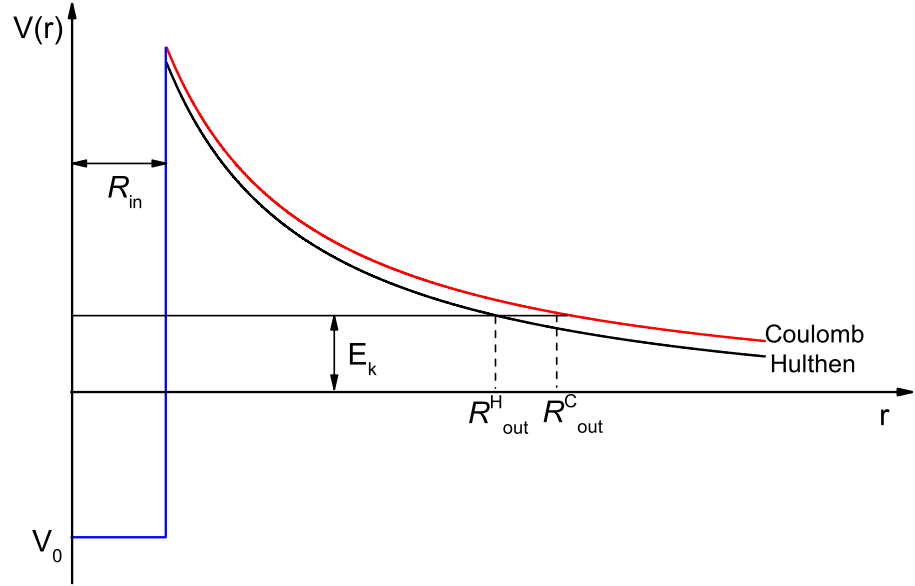}
\caption{(color online) The sketch map of total interaction potential between the emitted two protons and daughter nucleus
versus the center-of-mass distance of decay system. The external part of  potential barriers are represented by Coulomb and Hulth$\acute{\rm{e}}$n potential, respectively. }
\label{fig1}
\end{figure}

In the framework of Gamow-like model, for the total emitted two protons-daughter nucleus interaction potential $V(r)$, the inner potential associated with the nuclear interaction is represented by a square potential well, the outer electrostatic potential is represented by default of the Coulomb potential $V_{C}(r)$\cite{Liu2021}. It can be expressed as
\begin{equation}
\label{subeq:5}
V(r)=\left\{\begin{array}{llll}
-V_{0}&,\,\,\,0\leq r \leq R_{\rm{in}},\\
V_{C}(r)&,\,\,\,  r \textgreater R_{\rm{in}},
\end{array}\right.
\end {equation}
where $V_{0}$ is depth of the potential well. $V_{C}(r) = {Z_{2p} Z_{d} e^2} / {r}$ with $Z_{2p}$ and $Z_{d}$ being the proton numbers of the emitted two protons and the daughter nucleus, respectively. In this work, we modify the Gamow-like model by introducing an exponential-type electrostatic potential named Hulth$\acute{\rm{e}}$n potential $V_{H}(r)$ to investigate $2p$  radioactivity half-lives. As we can see from Fig. \ref{fig1}, this potential has the same behavior with the $V_{C}(r)$ at the short distance ($r\rightarrow0$) but drops exponentially more quickly at the long distance($r\gg0$), which was widely used in the fields of atomic, molecular and solid state physics, etc.\cite{Durand1981,Hall1985,Dutt1985,Lindhard1986,Pyykko1975,Olson1978} It can be defined as\cite{Hulthen1942,Hulthen1957}
\begin{equation}
V_{H}(r) = \frac{a Z_{2p} Z_{d} e^2}{e^{ar} - 1},
\label{subeq:6}
\end{equation}
 where $a$ is the screening parameter, which can determine the range of the potential, $i. e.$, the shortening of the exit radius.

In addition, we consisder the contribution of the centrifugal potential $V_{l}(r)$ on $2p$ radioactivity half-life in the modified Gamow-like model. It can be given by\cite{Morehead1995}
\begin{eqnarray}\label{subeq:7}
V_{l}(r)=\frac{ (l+{1}/{2})^2\hbar^2}{2\mu r^2},
\end{eqnarray}
where $\mu = m_{2p}m_{d} / (m_{2p} + m_{d}) \simeq 938.3 \times 2 \times A_{d} /A\,\rm{MeV}/c^2$ represents the reduced mass with $A_{d}$, $m_{2p}$ and $m_{d}$ being the mass number of daughter nucleus, the mass of the emitted two protons and the residual daughter nucleus, respectively. $l$ is the orbital angular momentum taken away by the emitted two protons, which satisfies the angular momentum and partity conservation law. Then, in the modified Gamow-like model, $V(r)$ can be written as
\begin{equation}
\label{subeq:8}
V(r)=\left\{\begin{array}{llll}
-V_{0}&,\,\,\,0\leq r \leq R_{\rm{in}},\\
V_{H}(r) + V_{l}(r)&,\,\,\,  r \textgreater R_{\rm{in}}.
\end{array}\right.
\end {equation}

\section{RESULTS AND DISCUSSION}
\begin{figure}
\centering
\includegraphics[width=15.0cm]{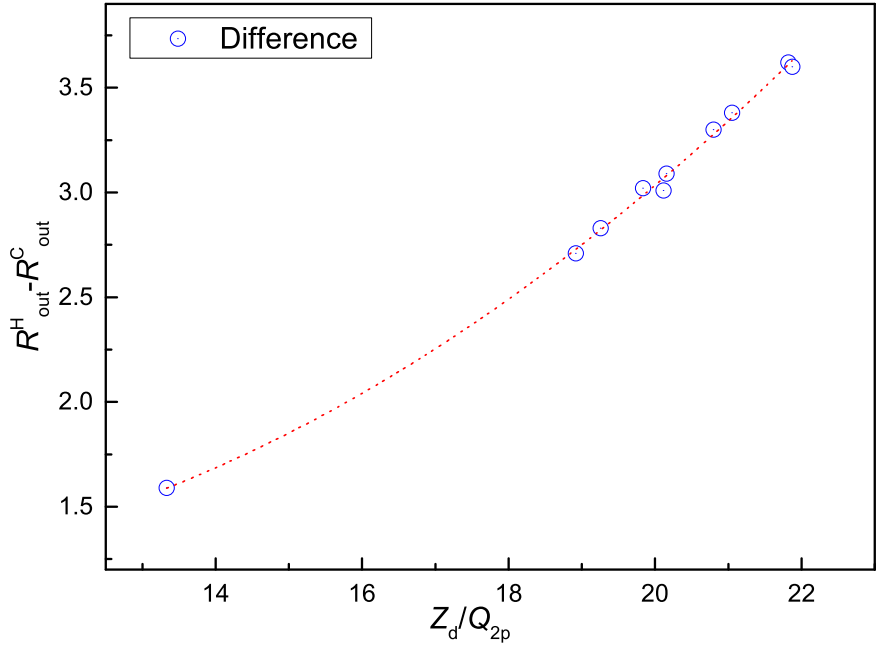}
\caption{(color online) The variation of $\emph{R}^{\emph{H}}_{\rm{out}}$$-$$\emph{R}^{\emph{C}}_{\rm{out}}$ with the increase of $Z_{d}/Q_{2p}$, while $R^{\emph{H}}_{\rm{out}}$ and $R^{\emph{C}}_{\rm{out}}$ are obtained by the modified Gamow-like model and the Gamow-like model, respectively.}
\label{fig2}
\end{figure}
Recently, a few of works indicated that the electrostatic shielding effect will affect the half-lives of unstable nuclei to a certain extent \cite{Jeppesen2007,Wan2015,Wan2016}. Meanwhile, the influence of this effect is correlated with the proton number of daugher nucleus and the decay energy. In 2019, based on the Gamow-like model consisdering the electrostatic shielding effect, we systematically studied the half-lives of $\alpha$ decay and proton radioactivity nuclei, respectively\cite{Cheng2019,Chen2019}. The calculated results are in better agreement with the experimental data for the nuclei with small proton number of daugher nucleus and large decay energy. 

\begin{figure}
\centering
\includegraphics[width=15.0cm]{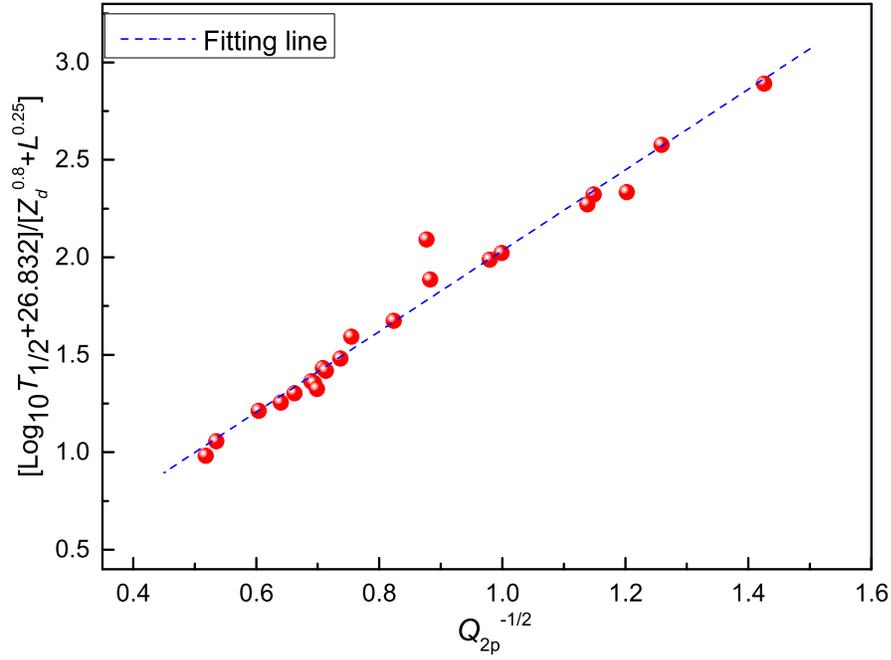}
\caption{(color online) The relationship between the quantity $[\rm{log_{10}}{\emph{T}}_{1/2} + 26.832]/(Z_\emph{{d}}^{0.8}+\emph{l}^{\,0.25})$ and $Q_{2p}^{-1/2}$ for the predictions obtained by modified Gamow-like model.}
\label{fig3}               
\end{figure}

Taking into account the continuous accumulations of experimental data in recent decades, to further explore the influence of electrostatic shielding effect on the $2p$ radioactivity half-lives of nuclei close to the proton drip line is an interesting topic. In this work, we calculate the half-lives of the true $2p$ radioactivity nuclei ($^{19}$Mg, $^{45}$Fe, $^{48}$Ni, $^{54}$Zn and $^{67}$Kr) by using the modified Gamow-like model. In this model, the two adjustable parameters $i. e. ,$ the effective nuclear radius parameter $r_{0}$ is set as 1.14 fm to unite with the value for proton radioactivity\cite{Chen2019}, the screening parameter $a$ = 1.808$\times10^{-3} \rm{\,fm}^{-1}$, which is determined by fitting the experimental data. In order to see the dependence of electrostatic shielding effect on the charge number $Z_{d}$ and released energy $Q_{2p}$ of $2p$ radioactivity nucleus. Fig. \ref{fig2} shows the variation of the differences in $\emph{R}_{\rm{out}}$ values for the pure Coulomb and Hulth$\acute{\rm{e}}$n potential ($\emph{R}^{\emph{H}}_{\rm{out}}$$-$$\emph{R}^{\emph{C}}_{\rm{out}}$) with $Z_{d}/Q_{2p}$. From this figure, it is obviously that $\emph{R}^{\emph{H}}_{\rm{out}}$$-$$\emph{R}^{\emph{C}}_{\rm{out}}$ monotonically increases as $Z_{d}/Q_{2p}$. This indicates that the electrostatic shielding effect is more obvious for nuclei with smaller $Q_{2p}$ and larger $Z_{d}$. The calculated results are listed in the last column of Table \ref{table1}. Meanwhile, the calculated $2p$ radioactivity half-lives using the Gamow-like model\cite{Liu2021}, effective liquid drop model (ELDM)\cite{Goncalves2017}, generalized liquid drop model (GLDM)\cite{Cui2020} and Sreeja formula\cite{Sreeja2019} are listed in the fifth to eighth columns for comparison, respectively. In this Table, the first four columns denote the $2p$ radioactivity parent nucleus, experimental $2p$ radioactivity released energy $Q_{2p}$ which are taken from the corresponding literatures, the angular momentum $l$ taken away by the emitted two protons and the  logarithmical experimental $2p$ radioactivity half-life $\rm{log}_{10}\emph{T}_{1/2}^{\rm{\,exp}}$, respectively. From this table we can find that the calculated results obtained by using the modified Gamow-like model can well reproduce the experimental data as well as the ones obtained by using Gamow-like model, ELDM, GLDM and Sreeja formula. To further indicate the good agreement between the experimental data and calculated ones, the deviation $\sigma$ = $\sqrt{\sum{(\rm{log}_{10}{\emph{T}_{1/2}^{\rm{\,calc}}}-\rm{log}_{10}{\emph{T}_{1/2}^{\rm{\,exp}}})^2}/n}$ are introduced. Based on the experimental data, we obtain the $\sigma$ values are equal to 0.825, 0.847, 0.531, 0.852 and 1.222 for the modified Gamow-like model, Gamow-like model, ELDM, GLDM and Sreeja formula, respectively. Because the electrostatic shielding effect is consisdered, the calculated half-lives obtained by using the modified Gamow-like model can much better reproduce the experimental data compared with the ones obtained by using the Gamow-like model. It indicates the particular importance of this effect. The present results are acceptable in view of the complexity of $2p$ radioactivity process.

In the following, using the modified Gamow-like model, we predict the half-lives of 22 possible $2p$ radioactivity candidates. The $2p$ radioactivity released energy $Q_{2p}$ of these candidates are extracted from the evaluated atomic mass table AME2016\cite{Huang2017,Wang2017}, the angular momentum $l$ taken away by the emitted two protons are determined by the angular momentum and parity conservation laws.  The predicted results are compared with the ones obtained by using Gamow-like model, ELDM, GLDM and Sreeja formula, all the detailed results are shown in Table \ref{table2}. In this table, the first three columns denote the $2p$ radioactivity candidate, $Q_{2p}$ and $l$, respectively. The fourth to eighth columns represent the predicted results obtained by using Gamow-like model, ELDM, GLDM, Sreeja formula and modified Gamow-like model, respectively. As shown in this table, the predictions using modified Gamow-like model are agree well with the ones using other models and formula. 

In 1911, Geiger and Nuttal found there is a phenomenological relationship between the $\alpha$ decay half-life $T_{1/2}^{\alpha}$ and $\alpha$ decay energy $Q_{\alpha}$\cite{Geiger1911},  which is so-called G-N law and expressed as
 \begin{equation}
\label{subeq:9}
\rm{log_{10}}{\emph{T}}_{1/2}^{\,\alpha} = \emph{a}\,\emph{Q}_{\alpha}^{\,-1/2} +\,\emph{b},
\end {equation}
where $\emph{a}$ and $\emph{b}$ are adjustable parameters.
 Recently, based on the G-N law, we put forward a two-parameter empirical formula for $2p$ radioactivity half-lives $T_{1/2}^{2p}$ by consisdering the contributions of the $Z_{d}$ and $l$ on $T_{1/2}^{2p}$\cite{Liu202102}. It can be expressed as
\begin{equation}
\label{subeq:10}
\rm{log_{10}}{\emph{T}}_{1/2}^{\,2\emph{p}} = 2.032\,({\emph{Z}_\emph{d}}^{\,0.8}+{\emph{l}}^{\,0.25})\,{\emph{Q}_{2\emph{p}}}^{-1/2} - 26.832.
\end {equation}
In order to test the reasonableness of the predicted results obtained by using modified Gamow-like model,  we plot the relationship between the quantity $[\rm{log_{10}}{\emph{T}}_{1/2} + 26.832]/(\emph{Z}_{\emph{d}}^{\,0.8}+\emph{l}^{\,0.25})$ and $Q_{2p}^{-1/2}$ in Fig. \ref{fig3}. As we can see from this figure, the $[\rm{log_{10}}{\emph{T}}_{1/2} + 26.832]/(\emph{Z}_{\emph{d}}^{\,0.8}+\emph{l}^{\,0.25})$ versus $Q_{2p}^{-1/2}$ exist an apparent linear behavior. We hope the present research can provide theoretical reference for the future experiments.

\begin{table*}[!hbt]
\centering
\caption{The comparisons between the logarithmic form of experimental $2p$ radioactivity half-lives $\rm{log}_{10}{\emph{T}}_{1/2}^{\;expt}$  and the logarithmic form of calculated $2p$ radioactivity ones by using five different theoretical models. The $\rm{log}_{10}{\emph{T}}_{1/2}^{\;expt}$ and experimental $2p$ radioactivity released energy $Q_{2p}$ are obtained from the corresponding references.}
\label{table1}
\renewcommand\arraystretch{1.5}
\setlength{\tabcolsep}{4pt}
\setlength\LTleft{-10in}
\setlength\LTright{-10in plus 1 fill}
\begin{tabular}{ccccccccc}
\hline\noalign{\smallskip}
\hline\noalign{\smallskip}
\multirow{2}{*}{Nuclei} & \multirow{2}{*}{$Q_{2p}$ (MeV)}&\multirow{2}{*}{$l$}& \multicolumn{6}{c}{$\rm{log}_{10}{\emph{T}}_{1/2}$(s)}\\
 \cmidrule(lr){4-9} 
 & & &Expt &Gamow-like&ELDM\cite{Goncalves2017}&GLDM\cite{Cui2020}&Sreeja\cite{Sreeja2019}&This work\\
\hline
$^{19}$Mg	&	0.750\cite{Mukha2007}	&0&$	-11.40 	$\cite{Mukha2007}&$	-11.46 	$&$	-11.72 	$&$	-11.79 	$&$	-10.66 	$&$-11.39$ 	\\
\hline
$^{45}$Fe	&	1.100\cite{Pfutzner2002}	&0&$	-2.40$\cite{Pfutzner2002}	&$	-2.09 	$&$-$&$	-2.23 	$&$	-1.25 	$&$-2.28$ 		\\
	&	1.140\cite{Giovinazzo2002}	&0&$	-2.07$\cite{Giovinazzo2002} 	&$	-2.58 	$&$-$&$	-2.71 	$&$	-1.66 	$&$-2.73$ 	\\
	&	1.154\cite{Dossat2005}	&0&$-2.55$\cite{Dossat2005}&$	-2.74 	$&$	-2.43 	$&$	-2.87 	$&$	-1.80 	$&$-2.88$		\\
	&	1.210\cite{Audirac2012}	&0&$-2.42$ \cite{Audirac2012}	&$	-3.37 	$&$	-	$&$	-3.50 	$&$	-2.34 	$&$-3.46$ 		\\
	\hline
$^{48}$Ni	&	1.290\cite{Pomorski2014}	&0&$-2.52$\cite{Pomorski2014}&$	-2.59 	$&$-$&$	-2.62 	$&$	-1.61 	$&$-2.69$ 	\\
	&	1.350\cite{Dossat2005}	&0&$	-2.08 	$\cite{Dossat2005}&$	-3.37 	$&$-$&$	-3.24 	$&$	-2.13 	$&$-3.27$		\\
	\hline
$^{54}$Zn	&	1.280\cite{Ascher2011}	&0&$	-2.76$\cite{Ascher2011}&$	-0.93 	$&$-$&$	-0.87 	$&$	-0.10 	$&$-1.12$		\\
	&	1.480\cite{Blank2005}	&0&$-2.43$\cite{Blank2005}&$	-3.01 	$&$	-2.52 	$&$	-2.95 	$&$	-1.83 	$&$-3.05$ 	\\
	\hline
$^{67}$Kr	&	1.690\cite{Goigoux2016}	&0&$	-1.70 	$\cite{Goigoux2016}&$	-0.76	$&$	-0.06 	$&$	-1.25 	$&$	0.31 	$&$-0.84$\\
\noalign{\smallskip}\hline
\noalign{\smallskip}\hline
\end{tabular}
\end{table*}

\begin{table*}[!hbt]
\centering
\caption{The comparisons of the logarithmic form of predicted half-lives of the possible $2p$ radioactivity candidates, whose $2p$ radioactivity is energetically allowed or observed but not yet quantified in NUBASE2016\cite{Audi2017}.}
\label{table2}
\renewcommand\arraystretch{1.5}
\setlength{\tabcolsep}{4pt}
\setlength\LTleft{-10in}
\setlength\LTright{-10in plus 1 fill}
\begin{tabular}{cccccccc}
\hline\noalign{\smallskip}
\hline\noalign{\smallskip}
 \multirow{2}{*}{Nuclei} & \multirow{2}{*}{$Q_{2p}$ (MeV)} & \multirow{2}{*}{$l$}&
   \multicolumn{5}{c}{$\rm{log}_{10}{\emph{T}}_{1/2}^{\,Pre}$(s)} \\
 \cmidrule(lr){4-8} 
 & & &Gamow-like\cite{Liu2021}&ELDM\cite{Goncalves2017}&GLDM\cite{Cui2020}&Sreeja\cite{Sreeja2019}&This work\\
\hline
$^{22}$Si	&	1.283	&	0	&$	-13.25 	$&$	-13.32 	$&$	-13.30 	$&$	-12.30 	$&$-13.06$	\\
$^{26}$S	&	1.755	&	0	&$	-13.92 	$&$	-13.86 	$&$	-14.59 	$&$	-12.71 	$&$-13.67$	\\
$^{34}$Ca	&	1.474	&	0	&$	-10.10 	$&$	-9.91 	$&$	-10.71 	$&$	-8.65 	$&$-9.92$	\\
$^{36}$Sc	&	1.993	&	0	&$	-12.00 	$&$	-11.74 $&$	-$&$	-10.30 	$&$-11.73$	\\
$^{38}$Ti	&	2.743	&	0	&$	-13.84 	$&$	-13.56 	$&$	-14.27 	$&$	-11.93 	$&$-13.51$	\\
$^{39}$Ti	&	0.758	&	0	&$	-0.91 	$&$	-0.81 	$&$	-1.34 	$&$	-0.28 	$&$-1.32$	\\
$^{40}$V	&	1.842	&	0	&$	-10.15 	$&$	-9.85 	$&$	-$&$	-8.46 	$&$-9.92$	\\
$^{42}$Cr	&	1.002	&	0	&$	-2.65 	$&$	-2.43 	$&$	-2.88 	$&$	-1.78 	$&$-2.85$	\\
$^{47}$Co	&	1.042	&	0	&$	-0.42 	$&$	-0.11 	$&$	-	$&$	0.21 	$&$-0.73$	\\
$^{49}$Ni	&	0.492	&	0	&$	14.54 	$&$	14.64 	$&$	14.46 	$&$	12.78 $&$12.34$	\\
$^{56}$Ga	&	2.443	&	0	&$	-8.57	$&$	-8.00 	$&$	-	$&$	-6.42 	$&$-8.28$	\\
$^{58}$Ge	&	3.732	&	0	&$	-12.32 	$&$	-11.74 	$&$	-13.10 	$&$	-9.53 	$&$-11.90$	\\
$^{59}$Ge	&	2.102	&	0	&$	-6.31 	$&$	-5.71 	$&$	-6.97 	$&$	-4.44 	$&$-6.11$	\\
$^{60}$Ge	&	0.631	&	0	&$	14.24 	$&$	14.62 	$&$	13.55 	$&$	12.40 	$&$12.33$	\\
$^{61}$As	&	2.282	&	0	&$	-6.76 	$&$	-6.12 	$&$	-	$&$	-4.74 	$&$-6.52$	\\
\hline
$^{10}$N	&	1.300	&	1	&$	-17.36 	$&$	-17.64 	$&$		-$&$	-20.04 	$&$-17.16$	\\
$^{28}$Cl	&	1.965	&	2	&$	-13.11 	$&$	-12.95 	$&$	-	$&$	-14.52 	$&$-12.77$	\\
$^{32}$K	&	2.077	&	2	&$	-12.49$&$	-12.25 	$&$	-	$&$	-13.46 	$&$-12.14$	\\
$^{57}$Ga	&	2.047	&	2	&$	-5.91 	$&$	-5.30 	$&$	-	$&$	-5.22 	$&$-5.66$	\\
$^{62}$As	&	0.692	&	2	&$	14.06 	$&$	14.52 	$&$	-	$&$	13.83 	$&$12.37$	\\
$^{52}$Cu	&	0.772	&	4	&$	8.94 	$&$	9.36 	$&$	-	$&$	8.62 	$&$8.10$	\\
$^{60}$As	&	3.492	&	4	&$	-9.40 	$&$	-8.68 	$&$	-	$&$	-10.84 	$&$-8.86$	\\
\noalign{\smallskip}\hline
\noalign{\smallskip}\hline
\end{tabular}
\end{table*}

\section{Summary}
In summary, consisdering the electrostatic screening effect on Coloumb potential together with the influence of centrifugal potential on $2p$ radioactivity half-life, the Gamow-like model proposed by Zdeb $et \ al.$ is modified. Using this modified model, we investigate the half-lives of true $2p$ radioactivity nuclei $^{19}$Mg, $^{45}$Fe, $^{48}$Ni, $^{54}$Zn and $^{67}$Kr. It is found that the calculated results are in good agreement with experimental data. In addition, using this modified Gamow-like model, we predict the half-lives of possible $2p$ radioactivity candidates. The predicted results are consistent with the ones obtained by using Gamow-like model, ELDM, GLDM and a four-parameter formula.

\section*{Acknowledgements}
This work is supported in part by the National Natural Science Foundation of China (Grants No. 11205083 and No.11975132), the Construct Program of the Key Discipline in Hunan Province, the Research Foundation of Education Bureau of Hunan Province, China (Grant No. 18A237), the Natural Science Foundation of Hunan Province, China (Grants No. 2018JJ2321), the Innovation Group of Nuclear and Particle Physics in USC, the Shandong Province Natural Science Foundation, China (Grant No. ZR2019YQ01) and the Opening Project of Cooperative Innovation Center for Nuclear Fuel Cycle Technology and Equipment, University of South China (Grant No. 2019KFZ10).

\appendix

\end{document}